\documentclass[prd,preprint,tightenlines,showpacs,showkeys,preprintnumbers,nofootinbib,amsmath,amssymb]{revtex4}
\usepackage[dvips,final]{graphicx}
  \usepackage{amssymb}
   \usepackage{amsmath}
    \usepackage{epsfig}
     \usepackage{bm}
      \usepackage{pifont}
\providecommand{\MSbar }{\ensuremath{ \overline{\rm MS} }}

\begin{document}
\preprint{\vbox{\hbox{RUB-TPII-01/09/2012}}}

\title{Taming Landau singularities in QCD perturbation
       theory:\\ The analytic approach 2.0\footnote{%
This is an extended and updated version of an invited plenary talk
at the International Conference \emph{Renormalization Group and Related
Topics (RG 2008)}, Dubna, Russia, September 1 - 5, 2008.}}

\author{N.~G.~Stefanis\footnote{E-mail:
        stefanis@tp2.ruhr-uni-bochum.de}}
\affiliation{Institut f\"{u}r Theoretische Physik II,
             Ruhr-Universit\"{a}t Bochum,\\
             D-44780 Bochum, Germany
             }

\date{\today}

\begin{abstract}
The aim of this topical article is to outline the fundamental ideas
underlying the recently developed Fractional Analytic Perturbation
Theory (FAPT) of QCD and present its main calculational tools together
with key applications.
For this, it is first necessary to review previous methods to
apply QCD perturbation theory at low spacelike momentum scales, where
the influence of the Landau singularities becomes inevitable.
Several concepts are considered and their limitations are pointed out.
The usefulness of FAPT is discussed in terms of two characteristic
hadronic quantities: the perturbatively calculable part of the pion's
electromagnetic form factor in the spacelike region and the Higgs-boson
decay into a $b\bar b$ pair in the timelike region.
In the first case, the focus is on the optimization of the prediction
with respect to the choice of the renormalization scheme and the
dependence on the renormalization and the factorization scales.
The second case serves to show that the application of FAPT to this
reaction reaches already at the four-loop level an accuracy of the
order of $1\%$, avoiding difficulties inherent in the standard
perturbative expansion.
The obtained results are compared with estimates from fixed-order and
contour-improved QCD perturbation theory.
Using the brand-new Higgs mass value of about 125~GeV, measured at the
Large Hadron Collider (CERN), a prediction for
$\Gamma_{H\to b\bar{b}}=2.4 \pm 0.15~{\rm MeV}$
is extracted.
\end{abstract}

\maketitle

\section{Introduction}
\label{sec:intro}

In writing this article, I have two main objectives.
The first is to chart the evolution of the analytic approach to
Quantum Chromodynamics (QCD) perturbation theory, focusing on the
key turning points in its theoretical development.
The second is to demonstrate how this approach works in terms of
selected applications.
Of course, I am forced to leave out many interesting applications
and aspects of the approach, which in turn means that my presentation
will not be complete.
I would refer the reader to the original references for further
reading and in order to study the subject in more technical detail.
The sorts of ideas and issues to be discussed in this presentation
include the following benchmarks:
\begin{itemize}
\item ``Ultraviolet (UV) freedom'', ``infrared (IR) slavery'', and
      Landau singularity.
\item Renormalizability, analyticity (in $Q^2$ and $g^2$), causality,
      and summability.
\item First remedies in the infrared: Color saturation, effective
      gluon mass.
\item Screening of the Landau singularities via Sudakov form factors.
\item Shirkov--Solovtsov analytic coupling---Euclidean and Minkowski
      space.
\item From a recipe to a paradigm: Analytic Perturbation Theory (APT).
\item More scales, more riddles: Logarithms of the factorization
      (evolution) scale---Naive and Maximal analytization.
\item Generalization of the analyticity requirement: From the coupling
      to the whole hadronic amplitude and its spectral density.
\item Creation of Fractional APT (FAPT) in the spacelike and timelike
      regions.
\item Summation of perturbation series, Sudakov gluon resummation
      (exponentiation vs. analytization),
      power corrections.
\item Crossing of heavy-flavor thresholds.
\end{itemize}

The article is organized as follows. In Sec.\ \ref{sec:UV-IR} I review
the two core characteristics of QCD: ultraviolet (UV) freedom and infrared
(IR) confinement.
The first attempts how to deal with the Landau singularity of the strong
running coupling in the spacelike momentum region are described in
Sec.\ \ref{sec:remedies}.
The analytization procedure of the coupling by Shirkov and Solovtsov is
presented in Sec.\ \ref{sec:analytization}, while its generalization to
hadronic observables in Sec.\ \ref{sec:paradigm}.
Further extensions to accommodate more than one large scale are discussed
in Sec.\ \ref{sec:riddles}, followed in Sec.\ \ref{sec:fractional} by the
analytization treatment of noninteger powers of the coupling that give rise
to fractional analytic perturbation theory (FAPT) in the Euclidean and
Minkowski space.
Applications of FAPT are presented in Sec.\ \ref{sec:APPL-pion} (pion's
electromagnetic form factor) and Sec.\ \ref{sec:higgs} (decay into a
bottom--antibottom quark pair of a scalar Higgs boson).
Conclusions are drawn in Sec.\ \ref{sec:concl}.

\section{Ultraviolet freedom and infrared confinement}
\label{sec:UV-IR}

Two key properties define QCD as the unbroken $SU(N_c=3)$
(with $N_c$ denoting the number of colors) Yang-Mills theory
describing strong interactions at the microscopic level: Asymptotic
freedom and confinement.
While the first property has been confirmed and was granted
a Nobel prize in Physics in the year 2004, the other is still a
matter of fact, with the underlying microscopic mechanism still at
large.
One claims that the reason for the non-observation of free quarks and
gluons in nature is due to their color content.
From this, one concludes that only color singlets are
observable---hence, defacto color confinement.

The main ingredient in justifying the UV freedom of QCD is that the
renormalized QCD coupling becomes small at large momenta:
$g^2 \to 0$ as $Q^2 \to \infty$ \cite{GW73,Pol73}.
This implies that though initially the coupling and the momentum scale
are independent parameters\footnote{Actually, there is no scale at all
inside the QCD Lagrangian---provided one ignores current quark
masses.
A scale dependence is generated through dimensional transmutation.}
they get linked to each other on account of the
renormalizability of QCD.
Hence, in the large-momentum region (or, equivalently, at short
distances), QCD can be applied perturbatively, in analogy to QED in
the sense of a weak-coupling expansion.
A conventional way to define the running strong coupling in the
$\MSbar$ scheme is given by
$(\alpha_s \equiv g^2/4\pi)$
\begin{equation}
  \alpha_s(\mu)
=
  \frac{12\pi}{\left(33-2N_{f}\right)
  \ln \left(\mu^{2}/\Lambda^{2}\right)}
  \left[
        1 - \frac{6(153-19N_{f})}{(33-2N_{f})^{2}} \
            \frac{\ln \left[\ln \left(\mu^{2}/\Lambda^{2}\right)
                      \right]}%
            {\ln \left(\mu^{2}/\Lambda^{2}\right)}
  \right]
  + \mathcal{O}\left[\frac{1}{\ln ^{3}\left(\mu^{2}/\Lambda^{2}\right)}
               \right] \ ,
\label{eq:alpha-s}
\end{equation}
which expresses it by means of the dimensional parameter
$\Lambda \equiv \Lambda_{\rm QCD}$, whereas $N_f$ represents the
number of active flavors.
For our considerations to follow it is more convenient to use the
abbreviation
$L \equiv \ln \left(\mu^{2}/\Lambda^{2}\right)$
and recast the above expression in terms of the first two coefficients
of the $\beta$ function
\begin{equation}
  \mu^2 \frac{d}{d\mu^2}\left(\frac{\alpha_s(\mu)}{4\pi}\right)
=
  \beta\left(\frac{\alpha_s(\mu)}{4\pi}\right) \ ,
\label{eq:beta}
\end{equation}
namely,
\begin{equation}
 b_0
=
 \frac{11}{3}C_{\rm A} - \frac{4}{3}T_{\rm R}N_{f}
=
 11 - 2N_{f}/3
\label{eq:b0}
\end{equation}
and
\begin{equation}
 b_1
=
   \frac{34}{3}C_{\rm A}^{2}
 - \left( 4C_{\rm F} + \frac{20}{3}C_{\rm A}\right)T_{\rm R}N_{f}
=
 102-38N_{f}/3 \ ,
\label{eq:1}
\end{equation}
where
$C_{\rm F}=(N_{c}^{2}-1)/2N_{c}=4/3$,
$C_{\rm A}=N_{c}=3$, and
$T_{\rm R}= 1/2$.
Then, Eq.\ (\ref{eq:alpha-s}) becomes
\begin{equation}
  \alpha_s[L(\mu)]
=
  \frac{4\pi}{b_0} \ \frac{1}{L}
  \left[
        1 - \frac{b_{1}}{b_{0}^{2}} \ \frac{\ln [L]}{L}
  \right]
  + \mathcal{O}\left[\frac{1}{L^{3}}\right] \ ,
\label{eq:alpha-s-compact}
\end{equation}
which is an approximate solution of the Renormalization-Group (RG)
equation\footnote{The first two coefficients $\beta_0$ and $\beta_1$
are scheme independent. The next coefficients $\beta_2$ and $\beta_3$
have also been analytically computed and confirmed by independent
calculations at the level of 3 and 4 loops (see \cite{KK09} for
references).}
\begin{equation}
  \frac{d}{dL} \left(\frac{\alpha_s(L)}{4\pi}
               \right)
=
  - b_0 \left(\frac{\alpha_s(L)}{4\pi}\right)^{2}
  - b_1 \left(\frac{\alpha_s(L)}{4\pi}\right)^{3} \ ,
\label{eq:ren-group-eq}
\end{equation}
exhibiting the asymptotic-freedom property, i.e.,
$\alpha_s \sim 1/\ln (\mu^{2}/\Lambda^{2}) \to 0$
as $\mu^{2}/\Lambda^{2}\rightarrow \infty$.
The scale parameter $\Lambda$ is not a generic parameter of the
QCD Lagrangian; its definition is arbitrary.
However, once defined, it becomes the fundamental constant of QCD
to be determined by experiment.

On the other hand, at $\mu^{2}=\Lambda^{2}$ the running strong
coupling has a Landau singularity that spoils analyticity.
At the one-loop level this singularity is a simple pole, whereas
at higher loops the corresponding Landau singularity has a more
complicated structure
(e.g., a square-root singularity at two loops, et cetera).
To restore analyticity in the infrared region
$0 \leq Q^2 \leq \Lambda^2$ and ensure causality in the whole $Q^2$
plane, one has to remove the Landau singularity at any loop order of
the perturbative expansion.
This may become a serious problem, both theoretically and
phenomenologically, for several hadronic processes---in particular
for exclusive processes like hadron form factors.
For example, in the case of the spacelike pion's electromagnetic form
factor---a typical exclusive process---most experimental data are
available in a momentum region, where repercussions of the Landau
singularity can influence (actually contaminate) the perturbative
calculation, as first shown and detailed in \cite{SSK99} and later
further analyzed in \cite{BPSS04}.
Another way to describe this behavior is to use the language of
information theory and consider a perturbatively calculable quantity,
like the scaled pion factor vs. the momentum transfer $Q^2$, as being
a \emph{dynamic range} of quality of the perturbative expansion.
In this context, the accuracy of the QCD perturbative result
(\emph{the signal}) depends on the magnitude of the running coupling.
The larger $Q^2$, the smaller the coupling and, hence, the higher the
precision of the (perturbative) calculation is.
As one approaches the Landau singularity, the coupling increases
rendering the perturbative expansion less and less precise, i.e.,
creating a kind of \emph{noise}.
When one reaches the Landau singularity, the ``noise''
(the uncertainty)
becomes overwhelming and one is unable to extract any information
(a clear ``signal'') from perturbation theory.
The next few sections will show in more detail how this problem can be
averted.

Of course, one may interpret the breakdown of perturbation theory at
low momenta $\sim \Lambda$, as a ``natural'' consequence of the
(unknown) confinement forces that prevail in this momentum region.
The problem is that this breakdown is not universal but depends on the
renormalization scheme adopted and the renormalization scale-fixing
procedure used.
For instance, adopting the Brodsky-Lepage-Mackenzie (BLM)
commensurate-scale procedure, when including higher-order QCD
perturbative corrections (see \cite{SSK99,BPSS04} for further details),
the BLM scale may become very low
(compared to the original momentum scale) and deeply intruding into the
``forbidden'' nonperturbative region, where the running coupling blows
up and perturbative QCD becomes invalid.
An example: Using the BLM procedure in the calculation of the pion's
electromagnetic form factor at NLO \cite{BPSS04}, the QCD running
coupling may still ``feel'' the Landau singularity at momenta as large
as $6$~GeV${}^2$, a region close to the largest momenta probed by
experiment at present for this reaction.
Hence, the (scheme-dependent) optimization of the QCD perturbative
expansion of typical hadronic quantities may depend in a crucial way
on the IR behavior of the running coupling.
It becomes clear that any attempt to apply an optimized
renormalization prescription (and scheme) in high fixed-order perturbative
calculations should make sure that the result is not distorted by
artificial logarithmic enhancements originating from the Landau
singularity.
Clearing away this obstruction is a real challenge.

\section{First remedies in the infrared}
\label{sec:remedies}

The simplest way to avoid the Landau singularity in Euclidean space is
to use a rigid IR cutoff on its value, say,
$\alpha_{s}^{\rm cutoff}\approx 0.5-0.7$,
such as to render the perturbative expansion sound.
A more sophisticated procedure is to assume that below some momentum
scale $\mu_{\rm IR}\gtrsim \Lambda$ color forces saturate,
so that quarks and gluons are confined within color-singlet states.
The IR cutoff scale $\mu_{\rm IR}$ can be conceived of as an effective
gluon mass $m_g$ generated by the dynamics of confinement, as proposed
by Cornwall (1982) \cite{Cor82} (for further references and applications,
see \cite{Ste99}).
This mechanism tames the Landau singularity, but the intrinsic scale
$m_g$ is an extraneous parameter to the approach and has to be
determined by other methods, e.g., on the lattice \cite{CoSo82}.
Indeed, the saturated coupling at one loop reads
\begin{equation}
\alpha_{s}^{\rm sat}(Q^2)
=
\frac{4\pi}{\beta_0
\ln[(Q^2+\lambda^2)/\Lambda^2]}
\label{eq:sat-alpha}
\end{equation}
with $\lambda^2=4m_g^2$ and is IR finite.
Here $m_g$ is a dynamical gluon mass with numerical values in the
range $(500\pm 200)$~MeV.
This scale my be thought of as being the average transverse momentum
of vacuum partons within the framework of nonlocal quark and gluon
condensates.
Technically, the role of the IR regulator $m_g$ is to ``freeze''
$\alpha_{s}(Q^{2})$ at low momenta $Q^2\lesssim \lambda^2$, so that the
coupling ceases to increase and flattens out.

An alternative approach to protect perturbatively calculated
observables was developed by Sterman and collaborators \cite{BS89,LS92},
in which $\alpha_{s}$ singularities were screened by Sudakov-type
damping exponentials ${\rm e}^{-S}$ that encapsulate the effects of
gluonic radiative corrections.
These are resumed contributions stemming from two-particle reducible
diagrams (giving rise to leading double logarithms), while two-particle
irreducible diagrams (entailing subleading single logarithms) are
absorbed into the hard-scattering part of the process.
Because ${\rm e}^{-S}$ drops to zero faster than any power of
$\ln\left(\mu^2/\Lambda_{\rm QCD}^{2}\right)$, it provides sufficient
IR protection against Landau singularities both for mesonic amplitudes
as well as for baryonic ones (see \cite{Ste99} for a review).
Indeed, in the axial gauge, all Sudakov contributions due to
unintegrated transverse momenta of the gluon propagators exponentiate
into suppressing Sudakov factors that suffice to render the Landau
singularity of the one-loop running coupling in the form factors of the
pion and the nucleon harmless, though the latter case is theoretically
much more involved---see \cite{BJKBS94,Ste95,Ste99}.
The main advantage of this approach relative to the previous one is
that it contains an intrinsic IR regulator in terms of the impact
parameter characterizing the inter-quark separations in the
hadrons.
Hence, there is no need for external IR regulators (like $m_g$)
because IR protection is provided \emph{in situ} by the Sudakov effect,
the underlying idea being that gluons with wavelengths larger than
the impact parameter ``see'' the quark configuration as a whole, i.e.,
in a color-singlet state and hence decouple.
It is worth noting that the Sudakov suppression can be simulated by a
saturated running coupling of the form of Eq.\ (\ref{eq:sat-alpha}), if
we assume that the IR cutoff parameter (i.e., the effective gluon mass)
has to be adjusted to each individual value of the varying external
momentum---for a mathematical version of this idea see \cite{Ste99}.
Let us close this discussion by remarking that the application of this
type of approach to the timelike region using analytic continuation is
not straightforward owing to the fact that not only the running
coupling has to be analytically continued but also the whole Sudakov
form factor.

A bridge here to what is to come in the next section with regard to
analytization.
In separate parallel developments, Krasnikov and Pivovarov (1982)
\cite{KP82}, and independently Radyushkin (1982) \cite{Rad82}, have
obtained analytic expressions for the one-loop running coupling
(and its powers) directly in Minkowski space using an integral
transformation from the spacelike to the timelike region reverse to
that for the Adler $D$-function.
In both approaches a resummed $\Lambda$-parametrization for
$\alpha_s(Q^2)$ in the spacelike region was employed in order to
construct in the timelike region an expansion for $R(q^2)$ in which all
$(\pi^2/L^2)^N$-terms ($L\equiv \ln s/\Lambda_{\rm QCD}^{2}$) are
summed explicitly.
This way, an analytic coupling in the timelike region was derived.
This sort of analytic coupling in Minkowski space was rediscovered
in the context of the resummation of fermion bubbles by Ball, Beneke,
and Braun (1995) \cite{BBB95}, partly in connection with techniques and
applications to the $\tau$ hadronic width \cite{BeJam08} (more references
are given in \cite{BMS06}).
Still other approaches to avoid ghost singularities in the running
coupling are referenced in \cite{SSK99,BMS06}.

\section{Into analytization}
\label{sec:analytization}

Shirkov and Solovtsov (1996) \cite{SS97} have invented an analytic
coupling based solely on Renormalization-Group (RG) invariance and
causality (spectrality) in terms of a dispersion relation, generalizing
the earlier work by Bogoliubov, Logunov, and Shirkov for QED \cite{BLS60}.
The key ingredient of their framework is the spectral density
$\rho_f(\sigma)$ with the aid of which an analytic running coupling
can be defined in the Euclidean region ($-q^2=Q^2$) using a
K\"{a}ll\'{e}n-Lehmann spectral representation:
\begin{eqnarray}
  \left[f(Q^2)\right]_\text{an}
   = \int_0^{\infty}\!
      \frac{\rho_f(\sigma)}
         {\sigma+Q^2-i\epsilon}\,
       d\sigma \ .
 \end{eqnarray}
The same spectral density $\rho_f(\sigma)= \textbf{Im}[f(-\sigma)]/\pi$
is then used to define the running coupling also in the timelike region
by means of a dispersion relation for the Adler function.

\begin{figure}[h]
\vspace{0.3cm}
\centerline{\includegraphics[width=0.38\textwidth]{
  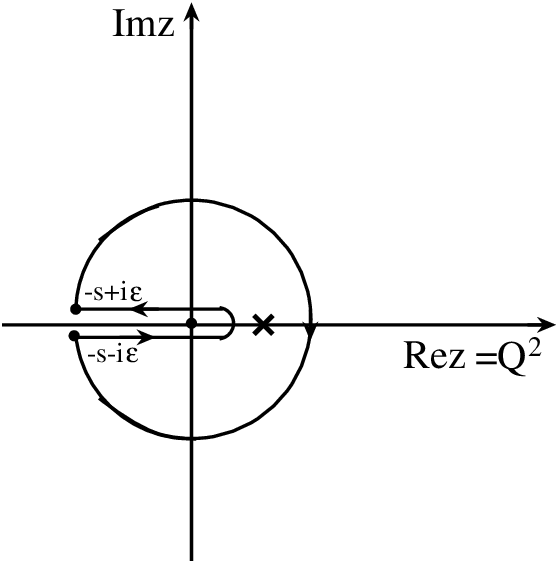}
  \vspace*{2mm}
~~~~~~~~~~~~~~\includegraphics[width=0.38\textwidth]{
 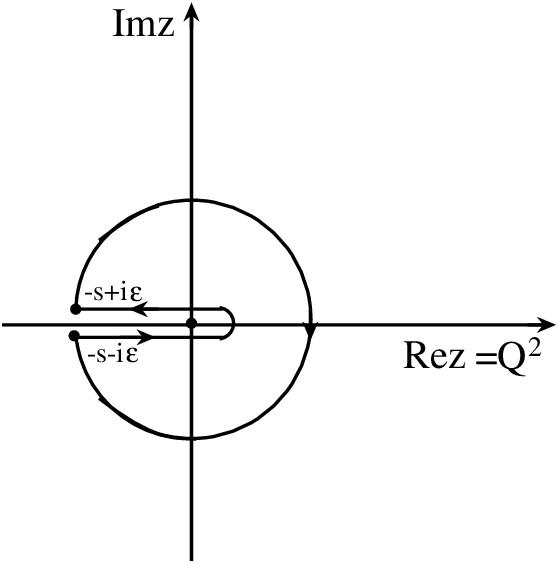}}
\vspace{0.2cm}
\caption{The Landau singularity, marked by the symbol
$\mbox{\boldmath$\times$}$ in the left panel, is absent by construction
in the analytic approach of Shirkov and Solovtsov \protect\cite{SS97}
(right panel).}
\label{fig:1}
\end{figure}

At the one-loop level, one obtains for the image of
$f(Q^2)=a(Q^2)\equiv (b_0/4\pi)\alpha_{s}(Q^2)$
in the Euclidean region \cite{SS97}
\begin{eqnarray}
  {\cal A}_1(Q^2)
&=&
  \int_{0}^{\infty} \! \frac{\rho(\sigma)}{\sigma+Q^2}\, d\sigma
=
  \frac{1}{L} - \frac{1}{e^L-1} \, ,
\label{eq:A_1}
\end{eqnarray}
while its counterpart in Minkowski space reads
\begin{eqnarray}
  {\mathfrak A}_1(s)
&=&
  \int_{s}^{\infty}\!\frac{\rho(\sigma)}{\sigma}\,d\sigma\
=
  \frac{1}{\pi}\,\arccos\frac{L_s}{\sqrt{\pi^2+L_s^2}} \, ,
\label{eq:U_1}
\end{eqnarray}
where
$
       L=\ln \left(Q^2/\Lambda_{\rm QCD}^2\right)
$
and
$
       L_s=\ln \left(s/\Lambda_{\rm QCD}^2\right).
$
Both couplings are analytic functions of their arguments.
The analyticity of ${\cal A}_1(Q^2)$ is ensured by the second
power-behaved term in (\ref{eq:A_1}) which removes the Landau (ghost)
pole, whereas ${\mathfrak A}_1(s)$ does not contain any singular term
at all.
This analytization concept was used earlier in QED by Redmond and
Uretsky (1958) \cite{RU58} in connection with the elimination of ghosts
in propagators, based on a term-by-term perturbation expansion in the
coupling constant.
The graphical representation of the removal of the Landau ghost in
the coupling is shown in Fig.\ \ref{fig:1}.
%

\section{The birth of APT. From a concept to a paradigm}
\label{sec:paradigm}

\begin{figure}[h]
\vspace{0.3cm}
\centerline{\includegraphics[width=0.45\textwidth]{
  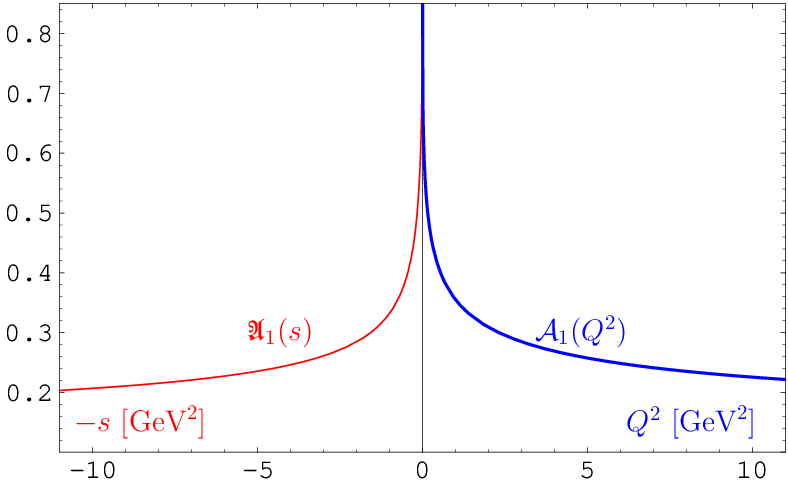}
  \vspace*{2mm}
~~~~~~\includegraphics[width=0.46\textwidth]{
 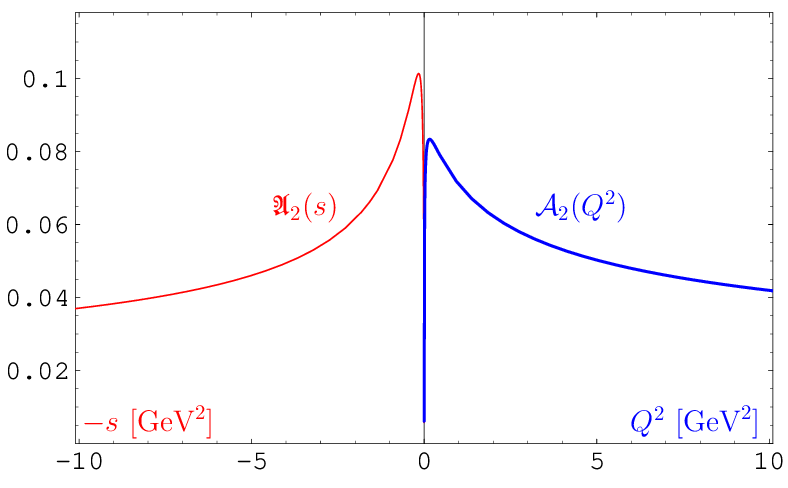}}
\vspace{0.2cm}
\caption{Analytic couplings in APT: spacelike
region---${\cal A}_{n}(Q^2)$ and
timelike region---${\mathfrak A}_{n}(s)$.
The left panel shows the first power (index), while the second
power (index) is given in the right panel.}
\label{fig:2}
\end{figure}

The simple analytization concept for the one-loop running coupling
in the spacelike region has been extended by its inventors and their
collaborators to higher loops and has been applied to several
characteristic hadronic processes.
Examples are the inclusive decay of a $\tau$-lepton into hadrons,
the momentum-scale and scheme dependence of the Bjorken and
the Gross-Llewellyn Smith sum rule, $\Upsilon$ decays into hadrons,
et cetera (the corresponding references can be found in \cite{BMS06}).
In Fig.\ \ref{fig:2}, we show the first power (left panel) and the
second power (right panel) of the analytic couplings in the Euclidean
(calligraphic notation) and in the Minkowski (Gothic notation) space.
All couplings are evaluated at the one-loop level.
%
One appreciates from this figure that the spacelike and timelike
analytic couplings have a distorted symmetry for finite values of
their arguments, though asymptotically,
i.e., for $L,L_s\to \infty$, they tend to the conventional coupling:
${\cal A}_{n}(L\to \infty) \Rightarrow {\mathfrak A}_{n}(L_s\to \infty)
\Rightarrow a^{n}(L\to \infty)~\mbox{for}~n\in \mathbb{N}$,
bearing in mind that $a=\left(b_{0}/4\pi\right)\alpha_s$.
As we will see later, this property is valid also for real indices
$\nu\in \mathbb{R}$.

Meanwhile, a systematic approach---a new paradigm for perturbative
calculations in QCD--- has emerged, termed Analytic Perturbation Theory
(APT).
In a nutshell, it can be summarized as follows.
The standard and the analytic couplings at one loop can be defined
\cite{BMS06} recursively for any integer power (index) $n\in\mathbb{N}$
by means of
\begin{equation}
\left(
\begin{array}{l}
  a^{n}(k)\\
{\cal A}_{n}(k)\\
 {\mathfrak A}_{n}(k)
 \end{array}
 \right)
 =
   \frac{1}{(n-1)!}\left( -\frac{d}{d k}\right)^{n-1}
\left(
\begin{array}{l}
  a^{1}(k)\\
{\cal A}_{1}(k)\\
 {\mathfrak A}_{1}(k)
 \end{array}
 \right),
\label{eq:generator}
\end{equation}
where
\begin{itemize}
\item $a^{n}$ (standard QCD with the scaled coupling
$a\equiv \alpha_s \frac{b_0}{4\pi} =1/L$,
$n\in \mathbb{N}$: power)
\item ${\cal A}_{n}$ (analytic in spacelike region,
$n\in \mathbb{N}$: index);
$k=L\equiv \ln(Q^2/\Lambda^2)$
\item ${\mathfrak A}_{n}$ (analytic in timelike region,
$n\in \mathbb{N}$: index);
$k= L_s\equiv\ln(s/\Lambda^2)$\ .
\end{itemize}
On the other hand, in terms of the spectral density
\begin{equation}
\rho_n(\sigma)
 =
 \frac{1}{\pi}\mathbf{Im}[a^n(-\sigma)]
 =
 \frac{\sin\left[n\arccos\left(L_s/\sqrt{L_s^2+\pi^2}\right)\right]}
           {n\,\pi\,\left[\sqrt{L_s^2+\pi ^2}\right]^{n}} \ ,
\label{eq:spec-dens}
\end{equation}
one has
\begin{subequations}
\begin{eqnarray}
  {\cal A}_{n}(Q^2)
& = &
  \int_{0}^{\infty}\frac{\rho_{n}(\sigma)}{\sigma+Q^2}d\sigma
= \frac{1}{L^n} - \frac{F(e^{-L},1-n)}{\Gamma(n)}
  \ ,\\
  {\mathfrak A}_{n}(s)
& = &
  \int_{s}^{\infty}\frac{\rho_{n}(\sigma)}{\sigma}d\sigma
=
\frac{\sin\left[(n-1)\arccos\left(L_s/\sqrt{L_s^2+\pi^2}\right)\right]}
           {(n-1)\,\pi\,\left[\sqrt{L_s^2+\pi ^2}\right]^{n-1}}
   \ ,
\label{eq:couplins-sd}
\end{eqnarray}
\end{subequations}
where $F(z,n)$ is the transcendental Lerch function serving in
(13a) as a Landau-pole remover \cite{BMS06}.
In APT, hadronic quantities defined (in the Minkowski region) by
$\oint \! f(z)R(z)dz$
with
\begin{equation}
  R^\text{PT}(z)
=
  \sum_{n} r_m \alpha_s^{m}(z)
  \stackrel{\rm APT}{\longrightarrow}
  \mathcal R^{\rm APT}(z)
=
  \sum_{n} d_m \mathfrak A_{m}(z)
\label{eq:had-qua}
\end{equation}
have no Landau singularities in the Euclidean region, because the
spacelike couplings ${\cal A}_n(z)$ are analytic functions.
One appreciates that APT transforms a power-series expansion into
a non-power-series, i.e., a functional, expansion.

Let us close this section by making some important remarks (for
references, see \cite{BMS06}):
\begin{itemize}
\item Analytization in Euclidean space means subtraction of the Landau
pole (at one loop).
\item Analytization in Minkowski space amounts to summation of $\pi^2$
terms.
\item Observables become non-power-series expansions
(i.e., n is an index not a power):
${\cal D}(Q^2)=\sum_{n}\,d_n\,{\cal A}_{n}(Q^2)$ (Euclidean space),
${\cal R}(s) = \sum_{n}\,d_n\,{\mathfrak A}^{}_{n}(s)$
(Minkowski space).
\item The elimination of ghost singularities in APT appears as a result
of causality (spectrality) and RG invariance, i.e., the pole remover is
not introduced by hand.
\item Two-loop expressions for the analytic couplings are possible by
means of the Lambert function (Magradze (2000)).
\item Still higher orders can be obtained with an approximate spectral
density and employing numerical integration (Shirkov (1999)).
\item Incorporation of heavy-quark flavors is possible
(``global APT''---Shirkov (2001)).
\item Modifications of APT in the deep IR are possible via a spectral
density which contains (nonperturbative) power corrections
(Alekseev (2006); Nesterenko and Papavassiliou (2005); Cveti\v{c} and
Valenzuela (2005)).
\end{itemize}

\section{More scales, more riddles}
\label{sec:riddles}

The above exposition suffices to demonstrate the usefulness of APT.
But where are its limitations?
Can the new paradigm be applied to more complicated cases that contain
more than one large momentum scale?
And can APT accommodate the crucial effect of evolution?
As we cannot sidestep these problems, we will have to dig deeper into
analytization.
This is not a terminal difficulty of the analytic approach---and we can
even benefit from it because doing so, we will ultimately expand APT to
noninteger powers of the coupling.
It is to these questions to which we focus now our attention.

To start with, we consider as a ``theoretical laboratory'' the
factorized part of the spacelike pion's electromagnetic form factor.
In NLO perturbation theory, this quantity can be written as a
convolution
$[A\otimes B \equiv \int_{0}^{1}dz A(z) B(z)]$
of a hard-scattering amplitude $T$ and two pion
distribution amplitudes,
$\varphi_{\pi}^{\rm in}$, $\varphi_{\pi}^{\rm out}$,
representing the incoming and outgoing pion bound states of quarks
with longitudinal momentum fractions $x$ and $y$, respectively.
$T$ is the amplitude for a collinear valence quark-antiquark pair,
struck by a highly virtual photon with the large momentum $Q^2$, and
contains hard-gluon exchanges characterized by the strong coupling
$\alpha_s$ to the considered order of the expansion.
Then, we have
\begin{equation}
  F_{\pi}^{\rm Fact}(Q^2)
=
  \varphi_{\pi}^{\rm in}(x,\mu_{\rm F}^2)\otimes
  T^{\rm NLO}_{\rm H}\left(x,y,Q^2;\mu_{\rm F}^2,\mu_{\rm R}^2\right)
                        \otimes
  \varphi_{\pi}^{\rm out}(y,\mu_{\rm F}^2)\ ,
\label{eq:F-fact}
\end{equation}
where
\begin{eqnarray}
 \varphi_\pi(x,\mu^2)
  = 6 x (1-x)
     \left[ 1
          + a_2(\mu^2) \, C_2^{3/2}(2 x -1)
          + a_4(\mu^2) \, C_4^{3/2}(2 x -1)
          + \ldots
     \right]
\label{eq:phi024mu0}
\end{eqnarray}
is the leading twist-2 pion distribution amplitude containing the
nonperturbative information on the pion quark structure in terms of the
Gegenbauer coefficients $a_n$ at the initial scale
$\mu^2 \approx 1$~GeV${}^{2}$, and $T$ is the sum of all quark-gluon
subprocesses up to the order $\alpha_{s}^{2}$ (its explicit form can be
found in the second paper in \cite{SSK99} and also in \cite{BPSS04}).
Note that $T$ depends explicitly on $\alpha_{s}$, whereas $\varphi_\pi$
has a more complicated dependence controlled by the evolution equation.
Besides, beyond the LO of perturbation theory, $T$ depends not only on
the large external momentum $Q^2$, but also on a second auxiliary
momentum scale $\mu_{\rm F}$ which separates the hard (perturbative)
from the soft (nonperturbative) dynamics on account of a factorization
theorem.
Invariably, the soft parts, i.e., the pion distribution amplitudes,
have to be evolved with appropriate anomalous dimensions from the
initial scale $\mu^2$ to the actual scale of observation $Q^2$.
As we will discuss in the next section, this additional scale has
influence on the analytization procedure that cannot be covered by
APT.

The first application of APT to $F_{\pi}^{\rm Fact}(Q^2)$ was
considered in \cite{SSK99}, in which the strong coupling and its powers
were replaced by their corresponding analytic images:
$a^n(L)\rightarrow \left({\cal A}_{1}(L)\right)^n$
(\emph{Naive analytization}
\cite{BPSS04}).
This treatment is strictly speaking incorrect because
$[{\cal A}_{1}(L)]^n\neq [a_s^n(L)]_{\rm An}$, but
phenomenologically it works rather good \cite{SSK99}.
An improvement of this approach was presented in \cite{BPSS04}, where
a \emph{Maximal analytization} of this quantity was performed.
This procedure associates to the powers of the running coupling their
own dispersive images, i.e.,
$[a_{s}^{n}(L)]_{\rm Max-An}={\cal A}_{n}(L)$.
It is worth emphasizing that the analytization treatment was applied
not only to the hard part $T$ but also to the anomalous dimensions,
governing the evolution of the pion distribution amplitudes, which can
be computed order-by-order in perturbation theory.
The differences between the two analytization procedures become
apparent by comparing the following two expressions:
\begin{eqnarray}
\emph{Naive~~Analytization}~~~~~~~~~~~~~~~~~~~~~\nonumber \\
 \left[Q^2 T_{\rm H}\left(x,y,Q^2;\mu_{\rm F}^2,\lambda_{\rm R} Q^2
                    \right)
 \right]_{\rm Naive-An}
\!\!& = &\!\!
  {\cal A}_{1}^{(2)}(\lambda_{\rm R} Q^2)\,
  t_{\rm H}^{(0)}(x,y)
\nonumber \\
&& \!\!\!\!\!\!\!\!\!\!\!\!\!\!\!\!\!\!\!\!\!\!\!\!\!\!\!\!\!\! +
  \frac{\left[{\cal A}_{1}^{(2)}(\lambda_{\rm R} Q^2)
        \right]^2}{4\pi}\,
   t_{\rm H}^{(1)}\left(x,y;\lambda_{\rm R},
                        \frac{\mu_{\rm F}^2}{Q^2}
                  \right)\\
\emph{Maximal~~Analytization}~~~~~~~~~~~~~~~~~\nonumber \\
  \left[Q^2 T_{\rm H}\left(x,y,Q^2;\mu_{\rm F}^2,\lambda_{\rm R} Q^2
                     \right)
  \right]_{\rm Max-An}
\!\! & = &\!\!
  {\cal A}_{1}^{(2)}(\lambda_{R} Q^2)\,t_{\rm H}^{(0)}(x,y)
\nonumber \\
&& \!\!\!\!\!\!\!\!\!\!\!\!\!\!\!\!\!\!\!\!\!\!\!\!\!\!\!\!\!\! +
  \frac{{\cal A}_{2}^{(2)}(\lambda_{\rm R} Q^2)}{4\pi}\,
  t_{\rm H}^{(1)}\left(x,y;\lambda_{\rm R},
                       \frac{\mu_{\rm F}^2}{Q^2}
                 \right) \ ,
\label{eq:naiv-vs-max}
\end{eqnarray}
where $\lambda_{\rm R} Q^2=\mu_{\rm R}^{2}$ serves as a
renormalization scale and $\mu_{\rm F}^2$ denotes the factorization
scale of the process.
The explicit form of the amplitudes $t_{\rm H}^{(0)}$ and
$t_{\rm H}^{(1)}$ is irrelevant for our arguments here; it can be found
in \cite{SSK99,BPSS04,BKS05}.
Furthermore, in \cite{SSK99}, the (naive) analytization procedure was
also applied after exponentiation to the Sudakov form factor in
next-to-leading logarithmic approximation of the cusp anomalous
dimension.

Some technical comments:
(a) The pole remover in the one-loop running coupling does not change
the leading double logarithmic behavior of the Sudakov factor, though
the cusp anomalous dimension in the Sudakov exponent receives
additional subleading contributions.
(b) Due to the absence of ghost singularities, the Sudakov exponents
are analytic functions of $Q^2$.
(c) The perturbatively calculable part of the pion's electromagnetic
form factor $Q^2F_{\pi}^{\rm Fact}(Q^2)$ receives the bulk of its
contributions from small transverse distances $b$ below
$b\cdot \Lambda_{\rm QCD}\lesssim 0.5$, i.e., it saturates in
exactly that region, where the application of QCD perturbation
theory is valid \cite{SSK99}.
However, the full treatment of the Sudakov-type gluon resummation in
impact space faces special subtleties which have not been fully
resolved until now.
First results within a toy model in momentum space
(see Appendix C of \cite{BMS05})
seem to indicate a conflict between analytization and exponentiation
of soft gluons.

\begin{figure}[t]
\vspace{0.3cm}
\centerline{\includegraphics[width=0.46\textwidth]{
  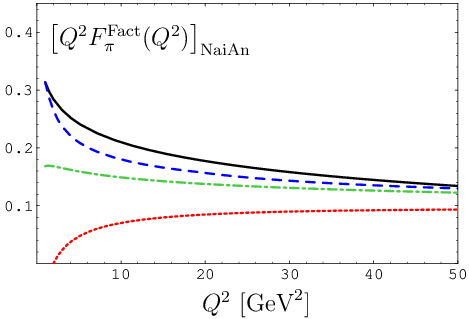}
  \vspace*{2mm}
~~~~~~\includegraphics[width=0.46\textwidth]{
 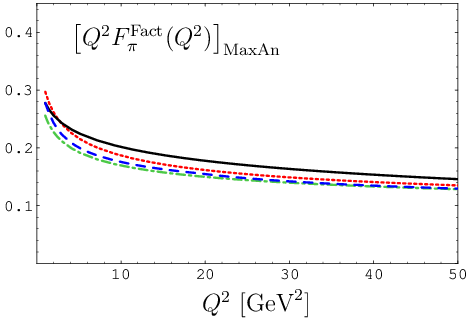}}
\vspace{0.2cm}
\caption{Results for $Q^2F_{\pi}^{\rm Fact}(Q^2)$ vs.\
    $Q^2$ with $\mu_{\rm R}^2=Q^2$, $\mu_{\rm F}^2=5.76$~GeV$^2$ and
    using Naive Analytization (left panel) and with Maximal
    Analytization (right panel) for various renormalization-scale
    settings described in \protect\cite{BPSS04}.
    All predictions have been calculated with the
    Bakulev-Mikhailov-Stefanis pion distribution amplitude
    (see \protect\cite{BPSS04} for references).}
\label{fig:3}
\end{figure}

Phenomenologically, already the ``naive'' analytization procedure
and even more the ``maximal'' application of APT have important
consequences:
(i) the artificial increase of the spacelike electromagnetic pion
form factor towards low $Q^2$ values---caused by the Landau
singularity---is absent,
(ii) the sensitivity of the results on the choice of the
renormalization scheme and the adopted scale setting is extremely
reduced relative to the standard power-series expansion of perturbative
QCD [Shirkov and Solovtsov (1998) and \cite{BPSS04}].
This behavior is shown graphically in Fig.\ \ref{fig:3} for the
scaled expression of $F_{\pi}^{\rm Fact}(Q^2)$ employing these two
analytization procedures and including leading-order evolution of
the pion distribution amplitude in APT.
%
The predictions shown were calculated for different
renormalization-scale settings, including the ``default'' choice,
$\mu_{\rm R}^{2}=Q^2$ (dashed line), the Brodsky-Lepage-Mackenzie
scale setting (dotted line), the so-called $\alpha_{\rm V}$ scheme
(dash-dotted line), et cetera, and employing the
Bakulev-Mikhailov-Stefanis pion distribution amplitude which has
$a_2(\mu_0^2=1~{\rm GeV}^2)=0.20$,
$a_4(\mu_0^2=1~{\rm GeV}^2)=-0.14$,
$a_n(\mu_0^2=1~{\rm GeV}^2)\approx 0~~(n > 2)$.
All these issues are discussed in detail in \cite{BPSS04}.
The main observation from this figure is that applying APT in its
maximal form to a typical exclusive quantity, like
$F_{\pi}^{\rm Fact}(Q^2)$,
not only provides IR stability, it also diminishes its renormalization
scheme and scale-setting dependence already at the NLO level.
Indeed, the predictions obtained in different schemes do not differ
significantly from each other in a wide range of momenta from low to
large $Q^2$ values.

To further appreciate the advantages of the analytic scheme relative
to the standard power-series expansion, bear in mind our statements in
Sec.\ \ref{sec:UV-IR} in the language of information theory.
Within this context, one may consider the variation of the
predictions---the uncertainty spread---obtained with different
renormalization schemes, as ``noise'', the prediction for
$F_{\pi}^{\rm Fact}(Q^2)$ itself being the ``signal''.
It becomes obvious from the right panel of Fig.\ \ref{fig:3} that
maximal analytization within APT provides a much better dynamical range
of quality because the signal-to-noise ratio is strongly enhanced in
the whole $Q^2$ range, especially at low $Q^2$.

\section{Fractionalizing APT}
\label{sec:fractional}

In the previous section we have discussed the advantages of APT
relative to the standard QCD perturbation theory and considered
a typical exclusive quantity, notably, the factorizable part of
the pion's electromagnetic form factor that exhibits a dependence
on potentially two large momentum scales: $Q^2$ and the
factorization scale $\mu_{\rm F}$ (that may also serve as the
evolution parameter).
Until now our discussion has not addressed the factorization-scale
dependence under the proviso of analytization.
But the considered example makes it obvious that, ultimately, one
has to accommodate the analytization of terms like
\begin{equation}
\begin{array}{lll}
  & \bullet~~~Z[L]={\rm e}^{\int^{a_s[L]}\frac{\gamma(a)}{\beta(a)}da}
  \to \left[a_s(L)\right]^{\gamma_0/2\beta_0}
  &  \longleftrightarrow  \, \, \,   {\rm RG~ at~ one~ loop} \\
  & \bullet~~~\left[a_s(L)\right]^{n}\ln[a_s(L)]
  &  \longleftrightarrow  \, \, \,   {\rm RG~ at~ two~ loops}\\
  & \bullet~~~\left[a_s(L)\right]^{n}L^m
  &  \longleftrightarrow  \, \, \,   {\rm Factorization}     \\
  & \bullet~~~\exp\left[-a_s(L) F(x)\right]
  &  \longleftrightarrow  \, \, \,
      {\rm Sudakov~ resummation}
  \
\end{array}
\end{equation}
typically appearing in perturbative calculations beyond LO, as in our
example.
Though such terms do not modify ghost singularities, say, the position
of the Landau pole in the one-loop approximation, they do contribute to
the spectral density.
To include such terms into the dispersion integral, one is actually
forced to modify the analyticity requirement and demand not only the
analyticity of the running coupling and its powers, but rather the
analyticity of the quark-gluon amplitude as a whole \cite{KS01}.
This generalized encompassing version of the analyticity requirement
demands that all terms that may contribute to the spectral density,
i.e., affect the discontinuity across the cut along the negative real
axis $-\infty<Q^2<0$, must be included into the analytization
procedure \cite{KS01,BMS05,BKS05}.
The use of this principle allows the inclusion into the dispersion
relation of logarithmic terms of the sort $\ln (Q^2/\mu_{\rm F}^2)$,
or products of such logarithms containing powers of the running coupling.
It turns out that precisely such terms are tantamount to
{\it fractional} (in fact, real) powers of the strong coupling,
pertaining to FAPT \cite{BMS05,BMS06}.

\begin{figure}[t]
\vspace{0.3cm}
\centerline{\includegraphics[width=0.26\textwidth]{%
 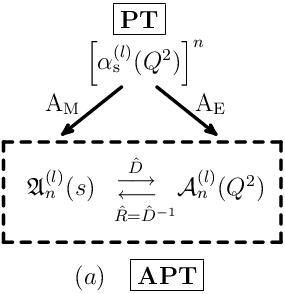}
 \vspace*{2mm}
 ~~~~~~\includegraphics[width=0.26\textwidth]{%
 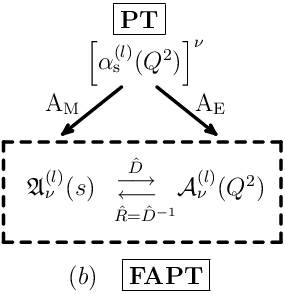}
 ~~~~~~\includegraphics[width=0.26\textwidth]{%
 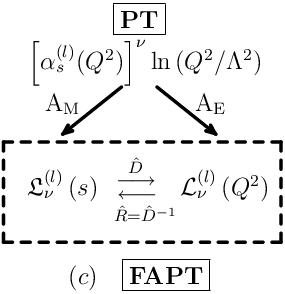}}
\vspace{0.2cm}
\caption{Implementation of analyticity in standard QCD perturbation
theory (PT), APT (a), and in FAPT (b) and (c).}
\label{fig:4}
\end{figure}

Before studying the consequences of this generalized analyticity
requirement for observables, let us first present the different
analytization concepts, we have discussed above, in mutual
comparison---Fig.\ \ref{fig:4}.
%
In this figure, the linear operations
$\textbf{A}_{\rm E}$ and ${\bf A}_{\rm M}$
define, respectively, the analytic running couplings in the Euclidean
(spacelike)
\begin{eqnarray}
 {\bf A}_{\rm E}\left[a^{n}_{(l)}\right]
  &=& {\cal A}^{(l)}_{n}
    ~~~{\rm with}~~~
   {\cal A}^{(l)}_{n}(Q^2)
  \equiv
   \int_0^{\infty}\!
    \frac{\rho^{(l)}_n(\sigma)}
         {\sigma+Q^2}\,
       d\sigma
\label{eq:A.E}
\end{eqnarray}
and the Minkowski (timelike) region
\begin{eqnarray}
 {\bf A}_{\rm M}\left[a^{n}_{(l)}\right]
  = {\mathfrak A}^{(l)}_{n}
  \text{~~~with~~~}
  {\mathfrak A}^{(l)}_{n}(s)
  \equiv
    \int_s^{\infty}\!
     \frac{\rho^{(l)}_n(\sigma)}
          {\sigma}\,
      d\sigma\ ,
\label{eq:A.M.rho}
\end{eqnarray}
where the loop order of the expansion is denoted by the superscript
$(l)$.
The above analytization operations can be represented by the
following two integral transformations from the timelike region
to the spacelike one
\begin{eqnarray}
  \hat{D}\big[{\mathfrak A}^{(l)}_{n}\big]
  &=&  {\cal A}^{(l)}_{n}
    ~~~{\rm with}~~~
   {\cal A}^{(l)}_{n}(Q^2)
  \equiv
   Q^2 \int_0^{\infty}\!
     \frac{{\mathfrak A}^{(l)}_{n}(\sigma)}
          {\big(\sigma+Q^2\big)^2}\,
      d\sigma
\label{eq:D-operation}
\end{eqnarray}
and for the inverse transformation
\begin{eqnarray}
  \hat{R}\big[{\cal A}^{(l)}_{n}\big]
  &=&
  {\mathfrak A}^{(l)}_{n}
  ~~~{\rm with}~~~
  {\mathfrak A}^{(l)}_{n}(s)
  \equiv
   \frac{1}{2\pi i}
   \int_{-s-i\varepsilon}^{-s+i\varepsilon}\!
    \frac{{\cal A}^{(l)}_{n}(\sigma)}
          {\sigma}\,
      d\sigma\, .
\label{eq:R-operation}
\end{eqnarray}
These two integral transformations are interrelated:
\begin{eqnarray}
 \label{eq:reciproc}
 \hat{D}\hat{R} = \hat{R}\hat{D} = 1
\end{eqnarray}
for the whole set of analytic images of the powers of the
coupling in the Euclidean, as well as in the Minkowski space,
$\big\{{\cal A}_n,{\mathfrak A}_n\big\}$, respectively, and at any
desired loop order of the perturbative expansion.
Then, Eq.\ (13a) generalizes from integer indices $n$ to any
real index $\nu\in \mathbb{R}$ to read (Euclidean space)
\begin{equation}
  {\cal A}_{\nu}(L)
      = \frac{1}{L^\nu}
      - \frac{F(e^{-L},1-\nu)}{\Gamma(\nu)}\, ,
\label{eq:lerch}
\end{equation}
where the first term corresponds to the conventional term of
perturbative QCD (with $L\equiv \ln(Q^2/\Lambda^2)$) and the second
one is entailed by the pole remover
(cf.\ $1/({\rm e}^{L}-1)$ at one loop).
The function ${\cal A}_{\nu}(L)$ is an entire function in the index
$\nu$ and has the following properties:
\begin{itemize}
\item ${\cal A}_{0}(L)=1$.
\item ${\cal A}_{-m}(L)=L^{m}=a^{-m}(L)$ for $m\in\mathbb{N}$.
Hence, the inverse powers of the analytic coupling and the original
running coupling coincide).
\item ${\cal A}_{m}(\pm\infty)=0$ for $m\geq 2$, $m\in\mathbb{N}$.
\item
$
 {\cal A }_{\nu}(L)
=
 -\left[1/\Gamma(\nu)\right]\sum_{r=0}^{\infty}\zeta(1-\nu-r)
                            \left[(-L)^{r}/r!\right]
$ for $|L|<2\pi$.
\end{itemize}
Note that the reduced transcendental Lerch function $F(z,\nu)$
is related to the Lerch function $\Phi(z,\nu, m)$ via
$z\Phi(z,\nu,1)\equiv F(z,\nu)$.
For a positive integer index, $\nu=m\geq2$, one has the relation
\begin{eqnarray}
  F(z,1-m)
=
  (-1)^{m}F\left(\frac1{z},1-m\right)\, ,
\label{eq:A3}
\end{eqnarray}
so that substituting Eq.\ (\ref{eq:A3}) in (\ref{eq:lerch}), one
arrives at
\begin{eqnarray}
 {\cal A}^{}_{m}(L)&=&(-1)^{m}{\cal A}^{}_{m}(-L)\ .
\label{eq:symm}
\end{eqnarray}

\begin{figure}[t]
\vspace{0.3cm}
\centerline{\includegraphics[width=0.45\textwidth]{%
 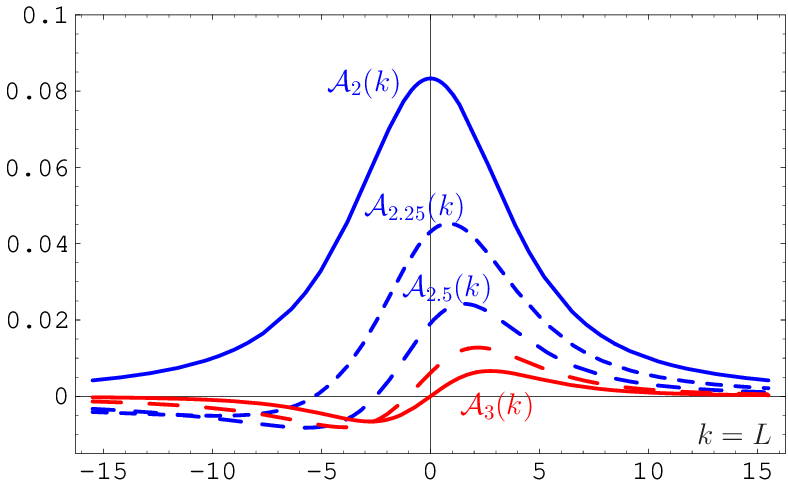}
 \vspace*{2mm}
 ~~~~~~\includegraphics[width=0.45\textwidth]{%
 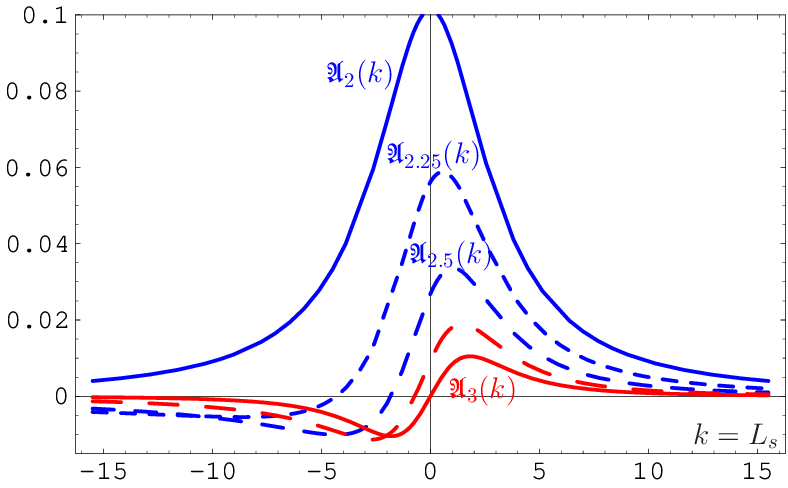}}
\vspace{0.2cm}
\caption{Comparison of the Euclidean (left panel) and the Minkowski
   (right panel) analytic couplings, ${\cal A}_{\nu}(k=L)$ and
   ${\mathfrak A}_{\nu}(k=L_s)$, respectively, for incremental
   changes of the index $\nu$ in the range 2 to 3.}
\label{fig:5}
\end{figure}

In Minkowski space, the analytic images of the running coupling
are completely determined by elementary functions \cite{BMS05}
($L_s\equiv \ln(s/\Lambda^2)$):
\begin{equation}
  {\mathfrak A}_{\nu}(L_s)
    = \frac{\sin\left[(\nu -1)
            \arccos\left(L_s/\sqrt{\pi^2+L_s^2}\right)
                \right]}
      {\pi(\nu -1) \left(\pi^2+L_s^2\right)^{(\nu-1)/2}}
\label{eq:gothic}
\end{equation}
from which, for example, we get
${\mathfrak A}_{0}(L_s)=1$, ${\mathfrak A}_{-1}(L_s)=L_s$;
${\mathfrak A}_{-2}(L)=L^2-\frac{\pi^2}{3}$;
${\mathfrak A}_{m}(L)=(-1)^{m}{\mathfrak A}_{m}(-L)$
for $ m\geq2$;
${\mathfrak A}_{m}(-\infty) = {\cal A}_{m}(-\infty) = \delta_{m,1}$;
${\mathfrak A}_{m}(\infty) = {\cal A}_{m}(\infty) = 0$
for $m\in \mathbb{N}$.
For the value $L=L_s=0$, the analytic couplings in both regions
are interrelated by the equation
\begin{eqnarray}
{\cal A}_{\nu}(0)
=
 \left[\frac{(\nu-1)\,\zeta(\nu)}{2^{\,\nu-1}}
 \right]
 {\mathfrak A}_{\nu}(0) \ ,
\end{eqnarray}
where
\begin{equation}
  {\mathfrak A}_{\nu}(0)
=
  \frac{\sin\left[(\nu-1)\pi/2\right]}{(\nu-1)\,\pi^{\nu}}
\label{eq:u-nu_0}
\end{equation}
and where $\zeta(\nu)$ is the Riemann $\zeta$ function, with the
coefficient in the bracket providing a quantitative measure
for the magnitude of the distortion of the ``mirror symmetry'',
mentioned before, for any $\nu \in \mathbb{R}$.
A graphical illustration of the analytic couplings ${\cal A}_{\nu}$
and ${\mathfrak A}_{\nu}$ is given in Fig.\ \ref{fig:5} which shows
the rate of change of these functions with respect to the index $\nu$
and the argument $L$ (or $L_s$).

%

The major characteristics of FAPT in comparison with the standard
perturbative expansion in QCD and the original APT  are collected in
Table \ref{Tab:char}, while further details can be found in
\cite{BMS05,BKS05,BMS06}.

 \begin{table}[t]
 \caption{FAPT versus standard QCD perturbation theory (SPT) and APT.}
  \begin{tabular}{|c|c|c|c|c|}\hline
  ~~~~~~Theory~~~~~~
           & ~~~~~~~SPT~~~~~~~
                     & ~~~~~APT~~~~~
                               & ~~~~~FAPT~~~~~         \\
  \hline\hline
  Space  & $\big\{a^\nu\big\}_{\nu\in\mathbb{R}}%
           \vphantom{^{\big|}_{\big|}}$
         & $\big\{{\cal A}_m,~{\mathfrak A}_{m}\big\}_{m\in\mathbb{N}}$
         & $\big\{{\cal A}_\nu,~{\mathfrak A}_{\nu}\big\}_{\nu\in\mathbb{R}}$
                                                        \\
  Series expansion
         & $\sum\limits_{m}f_m\,a^m(L)\vphantom{^{\big|}_{\big|}}$
         & $\sum\limits_{m}f_m\,{\cal A}_m(L)$
         & $\sum\limits_{m}f_m\,{\cal A}_m(L)$
                                                        \\
  Inverse powers
         & $\left[a(L)\right]^{-m}\vphantom{^{\big|}_{\big|}}$
         & ~{---}~
         & ${\cal A}_{-m}(L)=L^m$
                                                        \\
  Multiplication
         & $a^{\mu} a^{\nu}= a^{\mu+\nu}\vphantom{^{\big|}_{\big|}}$
         & ~{---}~
         & ~{---}~                                      \\
  Index derivative
         & $a^{\nu} \ln^{k}a\vphantom{^{\big|}_{\big|}}$
         & ~{---}~
         & $\frac{d^{k}}{d\nu^{k}}
         {{\cal A}_{\nu}^{} \choose {\mathfrak A}_{\nu}^{}}
          =\left[a^{\nu}\ln^{k}(a)\right]_{\rm an}$
                                                        \\
  \hline
\end{tabular}
\label{Tab:char}
\end{table}

Up to now we have actually concentrated our attention on the one-loop
approximation of the running coupling and its analytic images.
But the index derivative in FAPT gives us the possibility to obtain
two-loop expressions from those at one loop.
To achieve this goal, let us recall the two-loop RG equation for the
standard QCD normalized coupling
$a_{(2)}=b_{0}\alpha_s^{(2)}/4\pi$
\begin{eqnarray}
 \frac{d a_{(2)}(L)}{dL}
=
  - a_{(2)}^2(L)\left[1 + c_1\,a_{(2)}(L)\right]
  \quad \text{with}~c_1\equiv\frac{b_1}{b_0^2}
\label{eq:beta.new}
\end{eqnarray}
which assumes the following functional form
\begin{eqnarray}
  \frac{1}{a_{(2)}} +
  c_1
  \ln\left[\frac{a_{(2)}}{1+c_1 a_{(2)}}
     \right]
=
  L \ .
\label{eq:App-RGExact}
\end{eqnarray}
This equation has the following exact solution (Magradze 2000)
\begin{equation}
  a_{(2)}(L)
=
  - \frac{1}{c_1} \frac{1}{1 + W_{-1}(z_{W}(L))} \ ,
\label{eq:exact-Lambert}
\end{equation}
where
$
 z_{W}(L)
=
 (1/c_{1}) \exp{(-1 + i\pi - L/c_{1})}
$
with
$W_k$ $(k=0, \pm 1, \ldots)$
being the Lambert function, defined by
$
 z
=
 W(z)\exp{(W(z))}
$,
where its different branches are depicted by the index $k$.

With the above expression in our hands, we can immediately obtain
\cite{BMS05}
the spectral density at the two-loop level, $\rho_{\nu}^{(2)}(\sigma)$,
by virtue of Eq.\ (\ref{eq:spec-dens}), viz.,
\begin{equation}
  \rho_{\nu}^{(2)}(\sigma)
=
  \frac{1}{\pi} \mathbf{Im}[a_{(2)}^{\nu}(L-i\pi)]\ .
\label{eq:spec-dens-2-loop}
\end{equation}
Using this expression in Eqs.\ (\ref{eq:A.E}) and
(\ref{eq:A.M.rho}), we have
\begin{equation}
  {\cal A}_{\nu}^{(2)}(L)
=
  \int_0^{\infty}\!
    \frac{\rho^{(2)}_\nu(\sigma)}
         {\sigma+Q^2}\, d\sigma \, , ~~~~~
  {\mathfrak A}^{(2)}_{\nu}(s)
=
  \int_s^{\infty}\!
  \frac{\rho^{(2)}_\nu(\sigma)} {\sigma}\, d\sigma
\label{eq:A-2-loop}
\end{equation}
from which we derive $({\cal D}\equiv d^k/d\nu^k)$
\begin{eqnarray}
\label{eq:image-a2}
{\cal A }_{1}^{(2)}(L)
=
{\cal A }_{1}^{(1)}
        + c_1\,{\cal D}\,{\cal A }_{\nu=2}^{(1)}
        + c_1^2\,\left({\cal D}^{2}+{\cal D}-1\right)
          \,{\cal A }_{\nu=3}^{(1)}
        +{\cal O}\left({\cal D}^{3}\,{\cal A}_{\nu=4}^{(1)}\right) \, ,
\end{eqnarray}
where we have expanded $a_{(2)}(L)$ in terms of $a_{(1)}(L)$.
The above expression can be generalized to any real index $\nu$,
in both the Euclidean and the Minkowski space, to obtain
\begin{equation}
  \left(
        \begin{array}{l}
                        {\cal A }_{\nu}^{(2)}\\
                        {\mathfrak A}_{\nu}^{(2)}
        \end{array}
  \right)
 =
  \left(
        \begin{array}{l}
                        {\cal A }_{\nu}^{(1)}\\
                        {\mathfrak A}_{\nu}^{(1)}
        \end{array}
  \right)
 + c_1\,\nu\, {\cal D}\, \left(
                               \begin{array}{l}
                                               {\cal A }_{\nu+1}^{(1)}\\
                                               {\mathfrak A}_{\nu+1}^{(1)}
                               \end{array}
                         \right)
 + c_1^2\,\nu \left[\frac{\nu+1}{2!}\,{\cal D}^2\, + \,{\cal D}\, - \,1
              \right]
  \left(
        \begin{array}{l}
                        {\cal A }_{\nu+2}^{(1)}\\
                        {\mathfrak A}_{\nu+2}^{(1)}
        \end{array}
  \right) 
 + {\cal O}\left(c_1^{3}\right) \ .
\label{eq:a.2,u.2.nu}
\end{equation}
Though these expressions are not exact, the achieved accuracy is of the
order of about $1\%$ down to the value $L=L_s=0$ and, hence, they are
sufficient for most practical QCD applications.
A more accurate expression for the two-loop Minkowski coupling which
includes the contribution of the next higher order
${\cal O}\left({\cal D}^{4}\,{\mathfrak A}_{\nu + 4}^{(1)}\right)$
is given in \cite{BMS06}.
Moreover, one can find there the analogous three-loop expression with
a similarly good quality of convergence.

The particular advantage of FAPT is the possibility to perform the
analytization of powers of the running coupling multiplied by
logarithms of $L$ and its powers, e.g.,
$
 \left[\left(a_{(2)}(L)\right)^{\nu} L^{m}\right]_\text{an}
\equiv
 {\cal L}_{\nu,m}^{(2)}(L)
$.
As we emphasized before, such terms appear already in NLO calculations
of typical hadronic amplitudes, like form factors, or when taking into
account evolution effects.
Table \ref{tab:Feynman.Rules.FAPT} displays in approximate form the
analytization of such expressions.
Their usefulness will become evident below, when we will consider the
influence of such terms in the calculation of the factorized part of
the pion's electromagnetic form factor.

\begin{table}[t]
\caption{Calculational rules for FAPT
 with $L=\ln\left(Q^2/\Lambda^2\right)$,
 $m\in\mathbb{N}$, and $\nu\in\mathbb{R}$. $J_1$ is a Bessel function
 of the first kind and $\psi(z)=d\ln{\Gamma(z)}/dz$ denotes the
 Digamma function.
\label{tab:Feynman.Rules.FAPT}}
\begin{ruledtabular}
\begin{tabular}{l|l}
Standard QCD PT
             & QCD FAPT \\ \hline \hline
   $ a^\nu_{(1)}(L)=\frac{1}{L^\nu}\vphantom{^{\Big|}_{\Big|}}$
             &
   $\!\!\!\!\!\!\!\!\!\! {\cal A }_{\nu}^{(1)}(L)
   = \frac1{L^{\nu}}
   - \frac{F(e^{-L},1-\nu)}{\Gamma(\nu)}$
\\ 
$ a^\nu_{(l)}(L)\,\ln^m\left[a_{(l)}(L)\right]\vphantom{^{\Big|}_{\Big|}}$
           &
   $\!\!\!\!\!\!\!\!\!\! {\cal D}^{m}\,{\cal A }_{\nu}^{(l)}(L)
        \equiv\frac{d^m}{d\nu^m}\,
         \left[{\cal A }_{\nu}^{(l)}(L)\right]$
\\ 
  $ a^\nu_{(2)}(L) \vphantom{^{\Big|}_{\big|}}$
             &
    $\!\!\!\!\!\!\!\!\!\! {\cal A}_{\nu}^{(2)}(L)={\cal A }_{\nu}^{(1)}(L)
          + c_1\nu\,{\cal D}\,{\cal A }_{\nu+1}^{(1)}(L)+
          c_1^2\,\nu\left[\frac{(\nu+1)}{2}{\cal D}^{2}+{\cal D}-1\right]
          \!{\cal A }_{\nu+2}^{(1)}(L)$
          \\
          & ~~~~~~~~~~~~~~~$\!\!\!\!\!\!\!\!\!\!
          +~{\cal O}\left({\cal D}^{3}\,{\cal A}_{\nu+3}^{(1)}\right)
           \vphantom{_{\Big|}}$
\\ 
   $ a^\nu_{(2)}(L)\,L \vphantom{^{\Big|}_{\big|}}$
             &
   $\!\!\!\!\!\!\!\!\!\! {\cal L}_{\nu,1}^{(2)}(L)
          = {\cal A}_{\nu-1}^{(2)}(L)
          + c_1\,{\cal D}\,{\cal A }_{\nu}^{(2)}(L)
          + {\cal O}\left({\cal D}^{2}\,{\cal A}_{\nu+1}^{(2)}\right)
          $
\\           &
   $~~~~~ \approx {\cal A }_{\nu-1}^{(1)}(L)
          - c_1\nu\left[\frac{\ln(L)-\psi(\nu)}{L^{\nu}}
          + \psi(\nu){\cal A}_{\nu}^{(1)}(L)
          + \frac{{\cal D}F(e^{-L},1-\nu)}{\Gamma(\nu)}\right]
          \vphantom{_{\Big|}}$
\\ 
   $ a^\nu_{(2)}(L)\,L^2 \vphantom{^{\Big|}_{\big|}}$
             &
   $\!\!\!\!\!\!\!\!\!\! {\cal L}_{\nu,2}^{(2)}(L)
          = {\cal A }_{\nu-2}^{(2)}(L)
          + 2\,c_1\,{\cal D}\,{\cal A }_{\nu-1}^{(2)}(L)
          + c_1^2\,{\cal D}^{2}\,{\cal A }_{\nu}^{(2)}(L)
          - 2 c_1^2\,{\cal A}_{\nu}^{(2)}(L)
          $
\\          &
   $  ~~~~~~~~~~~~~
          \!\!\!\!\!\!\!\!\!\!
          +{\cal O}\left({\cal D}\,{\cal A}_{\nu+1}^{(2)}\right)
          $
\\          &
   $~~~~~ \approx {\cal A}_{\nu-2}^{(1)}(L)
         + c_1\,\nu\,{\cal D}\,{\cal A}_{\nu-1}^{(1)}(L)
         + c_1^2\,\left[\frac{\nu^2-\nu+4}{2}\right]\,
            {\cal D}^{2}\,{\cal A}_{\nu}^{(1)}(L) \vphantom{_{\Big|}}$
\\ 
$\exp\left[-x a(L)\right]\vphantom{^{\Big|}_{\Big|}}$
            &
$\!\!\!\!\!\!\!\!\!\! e^{-x/L}
+\sqrt{x}\sum_{m=1}e^{- mL} \frac{J_1(2 \sqrt{x m})}{\sqrt{m}}
\vphantom{_{\Big|}}~~\mbox{for}~L>0$
\end{tabular}
\end{ruledtabular}
\end{table}

\section{Spacelike pion's electromagnetic form factor in FAPT}
\label{sec:APPL-pion}

Having set up the calculational tools of FAPT, we now
detail concrete applications, starting with an example in the
spacelike region.
Consider again Eq.\ (\ref{eq:naiv-vs-max})---this time including into
the analytization procedure also the logarithmic term (labeled by the
acronym KS \cite{KS01} in order to indicate the analytization of the
amplitude as a whole):
$
  \ln(Q^2/\mu_{\rm F}^2)
=
  \ln (\lambda_\text{R} Q^2/\Lambda^2)
 -
  \ln (\lambda_\text{R}\mu_{\rm F}^2/\Lambda^2).
$
Then, we find ($\bar x = 1-x$)
\begin{eqnarray}
 \left[Q^2 T_\text{H}(x,y,Q^2;\mu_{\rm F}^2,\lambda_{\rm R} Q^2)
 \right]_{\rm KS}^{\rm An}
&  = &
  {\cal A}_{1}^{(2)}(\lambda_{\rm R} Q^2)\, t_{\rm H}^{(0)}(x,y)
+
   \frac{{\cal A}_{2}^{(2)}(\lambda_{\rm R} Q^2)}{4\pi}\,
   t_{\rm H}^{(1)}\left(x,y;\lambda_{\rm R},
  \frac{\mu_{\rm F}^2}{Q^2}\right)
\nonumber\\
&&
    +\ \frac{\Delta_{2}^{(2)}
  \left(\lambda_{\rm R} Q^2\right)}{4\pi}\,
      \left[C_{\rm F}\, t_{\rm H}^{(0)}(x,y)  \,
             \left(6 + 2 \ln(\bar{x}\bar{y})\right)
      \right] \ ,
\label{eq:TH-KS-6}
\end{eqnarray}
with
\begin{eqnarray}
   \Delta_{2}^{(2)}\left(Q^2\right)
&\equiv&
   {\cal L}_{2}^{(2)}\left(Q^2\right)
  -{\cal A}_{2}^{(2)}\left(Q^2\right)\,\ln\left[Q^2/\Lambda^2\right]
\label{eq:delta2-2}
\end{eqnarray}
encoding the deviation from the result obtained with the Maximal
Analytization procedure and where
\begin{eqnarray}
 {\cal L}_{2}^{(2)}\left(Q^2\right)
& \!\!\! \equiv \! \!\!\! &
   \left[\left(\alpha_{s}^{(2)}\left(Q^2\right)\right)^2
         \ln\left(\frac{Q^2}{\Lambda^2}\right)
   \right]_\text{KS}^{\rm An}
\!  = \! \frac{4\pi}{b_0}
     \left[\frac{\left(\alpha_{s}^{(2)}\left(Q^2\right)\right)^2}
                 {\alpha_{s}^{(1)}\left(Q^2\right)}
      \right]_\text{KS}^{\rm An}\ .
\label{eq:Log_Alpha_2_KS}
\end{eqnarray}
Carrying out the KS analytization, we get the approximate expression
\begin{eqnarray}
  {\cal L}_{2,1}^{{\rm appr}(2)}\left(Q^2\right)
   = \frac{4\pi}{b_0}\,
      \left[{\cal A}_{1}^{(2)}\left(Q^2\right)
      + c_1\,\frac{4\pi}{b_0}\,f_{\cal L}\left(Q^2\right)
      \right]
\label{eq:Log_Alpha_2_BMKS}
\end{eqnarray}
(which can be further improved by substituting the $c_1$-term in
Eq.\ (\ref{eq:Log_Alpha_2_BMKS}) by its two-loop expression
:
$
 {\cal L}_{2,1}^{\text{imp};(2)}[L]
=
 {\cal A}_{1}^{(2)}[L]
 + c_1\,{\cal D}\,
 {\cal A}_{2}^{(2)}[L]
$
---see \cite{AB08gfapt}), where
[cf.\ Table \ref{tab:Feynman.Rules.FAPT}, $a_{(2)}^{\nu=2}(L)L$]
\begin{eqnarray}
  f_{\cal L}\left(Q^2\right)
   = \sum_{n\geq0}
      \left[\psi(2)\zeta(-n-1)-\frac{d\zeta(-n-1)}{dn}\right]\,
       \frac{\left[-\ln\left(Q^2/\Lambda^2\right)
             \right]^n}{\Gamma(n+1)}
\label{eq:f_MS}
\end{eqnarray}
and $\zeta(z)$ is the Riemann Zeta-function.
The main effect of including the logarithmic term into the
analytization procedure of $F_{\pi}^{\rm Fact}(Q^2)$ is to suppress its
sensitivity on the variation of the factorization scale to a minimum.
It was shown in \cite{BKS05} that varying the factorization scale from
1~GeV${}^2$ to 10~GeV${}^2$, the form factor changes by a mere 1.5
percent reaching for a (hypothetical) factorization scale of
50~GeV${}^{2}$ just the level of about 2.5 percent.
Note here that it is possible to treat this problem in another way,
namely, to fix $\mu^2=Q^2$ from the very beginning and then use FAPT
to correctly account for the evolution factors
$E_n^\text{LO}(Q^2)\sim \alpha_s^{\nu_n}(Q^2)$
in the Gegenbauer coefficients $a_n(Q^2)$,
with $\nu_n$ being the corresponding fractional evolution exponents.
This type of calculation was performed in~\cite{AB08gfapt} and it was
shown that the difference between the two approaches, i.e.,
FAPT with fixed $\mu^2=5.76$ GeV$^2$
and
FAPT with the renormalization scale fixed at $\mu^2=Q^2$,
is of the order of 1.5\%.

To conclude, using FAPT, the dependence of a perturbatively
calculated QCD quantity on all perturbative scheme and scale settings
is diminished already at the NLO level.

\section{Higgs boson decay into a $b\bar{b}$ pair using FAPT in
Minkoeski space}
\label{sec:higgs}

The second FAPT application deals with the Higgs boson decay
into a $b\bar{b}$ pair in the timelike region.
We will present here only the main results and refer for the technical
details to \cite{BMS06,BMS10}.
A dedicated analysis of the uncertainties involved in the calculation
of the decay width of the Higgs boson into bottom quarks was recently
given by Kataev and Kim \cite{KK09}.

Consider now the decay of a scalar Higgs boson to a $b\bar{b}$ pair in
terms of the quantity $R_{\rm S}$ at the four-loop level from which one
can then obtain the width $\Gamma(\rm H\to b\bar{b})$.
Note that the application of FAPT in Minkowski space is not plagued by
Landau singularities.
However, the analytic continuation from the spacelike to the timelike
region induces so-called ``kinematical'' $\pi^2$ terms that may be
comparable in magnitude with the higher-order expansion coefficients
and have, therefore, to be summed.
The first step in our approach is the correlator of two scalar currents
$J^{\rm S}_b=\bar{\Psi}_b\Psi_b$
for bottom quarks with mass $m_b$, coupled to the scalar Higgs boson
with mass $M_{\rm H}$, and where $Q^2 = - q^2$, i.e.,
\begin{equation}
  \Pi(Q^2)
=
  (4\pi)^2 i\int dx {\rm e}^{iq \cdot x}\langle 0|\;T[\;J^{\rm S}_b(x)
  J^{\rm S}_{b}(0)\,]\;|0\rangle \ .
\label{eq:correlator}
\end{equation}
Then,
$
  R_{\rm S}(s)
 =
  \textbf{Im}\, \Pi(-s-i\epsilon)/{(2\pi\,  s)}
$
and one can express the decay width in terms of $R_{\rm S}$:
\begin{eqnarray}
  \Gamma({\rm H} \to b\bar{b})
= \frac{G_{\rm F}}{4\sqrt{2}\pi}M_{\rm H}
  m_{b}^{2}(M_{\rm H}) R_{\rm S}(s = M_{\rm H}^2) \ .
\label{decay_rate_for_b}
\end{eqnarray}
The quantity $R_{\rm S}$ is obtained from the Adler function $D$ via
analytic continuation into the Euclidean space using the transformation
${\bf A}_{\rm M}$ (or, equivalently, the integral transformation
$\hat R$), as shown in Fig.\ \ref{fig:4}.
The reason for this detour is that in the Euclidean region perturbation
theory works.
Hence, we write
\begin{eqnarray}
 \widetilde{D}_{\text{S}}(Q^2;\mu^2)
  &=& 3\,m_b^2(Q^2)
       \left[1+\sum_{n \geq 1} d_n(Q^2/\mu^2)~a_{s}^n(\mu^2)
        \right]\ .
\label{eq:D-s}
\end{eqnarray}
The second step is to calculate
\begin{eqnarray}
  \widetilde{R}_\text{S}(s)
\equiv
  \widetilde{R}_\text{S}(Q^2=s,\mu^2=s)
=
  3m^{2}_{b}(s)\bigl[ 1 + \sum_{n\geq 1}^{} r_{n}~a_{s}^n(s)
               \bigr]
  \label{eq:R-s}
\end{eqnarray}
using the techniques developed by Gorishnii et al. (1990)
\cite{GKLS90} and Chetyrkin et al. (1997) \cite{ChKS97}.

As already mentioned, the coefficients $r_n$ contain characteristic
$\pi^2$ terms originating from the integral transformation $\hat{R}$
of the powers of the logarithms entering $\widetilde{D}_{\rm S}$.
The latter is related to $\widetilde{R}_{\rm S}(s,s)$ by means of a
dispersion relation.
Notice that these logarithms have two different sources: one is the
running of $a_s$ in $\widetilde{D}_{\rm S}$, while the other is
related to the evolution of the heavy-quark mass $m_{b}^2(Q^2)$.
As a result, the coefficients $r_n$ in (\ref{eq:R-s}) are connected to
the coefficients $d_n$ in $\widetilde{D}_{\rm S}$ (calculable in
Euclidean space) and to a combination of the mass anomalous dimension
$\gamma_i$ and the $\beta$-function coefficients $b_j$ multiplied by
$\pi^2$ powers.
Appealing to the explicit results given in \cite{BMS06}, we remark
that the total amount of these terms is of the order of the
coefficient $d_4$.
FAPT can by construction account for these terms to all orders of
the perturbative expansion.
This is because in FAPT one does not need to expand the
renormalization factors into a truncated series of logarithms;
instead one can transform them by means of the
$\textbf{A}_\text{M}$-operation ``as a whole''.

The running mass in the $l$-loop approximation, $m_{(l)}$, can be
cast in terms of the renormalization-group invariant quantity
$\hat{m}_{(l)}$ to read
\begin{eqnarray}
  m_{(l)}^2(Q^2)
&=& \hat{m}_{(l)}^2
  \left[a_{s}(Q^2)\right]^{\nu_0}
  f_{(l)}(a_s(Q^2))\ ,
\label{eq:m2-hat-run}
\end{eqnarray}
where $\nu_0=2\gamma_0/b_0$ and the expansion of $f_{(l)}(x)$
at the three-loop order is given by
\begin{eqnarray}
  f_{(l)}(a_s)
&=&
  1 +  a_s\,\frac{b_1}{2b_0}\left(\frac{\gamma_1}{b_1}
    - \frac{\gamma_0}{b_0}\right)
    +  a_s^2\,\frac{b_1^2}{16\,b_0^2}
            \left[\frac{\gamma_0}{b_0}-\frac{\gamma_1}{b_1}
                 + 2\,\left(\frac{\gamma_0}{b_0}
                 -\frac{\gamma_1}{b_1}\right)^2
\right. \nonumber \\
&& \left.\!\!\!  + \frac{b_0 b_2}{b_1^2}\left(\frac{\gamma_2}{b_2}
                 -\frac{\gamma_0}{b_0}\right)
           \right]
    + O\left(a_s^3\right)\ .
\label{varphi}
\end{eqnarray}
Expanding the running mass in a power series according to
\begin{eqnarray}
  m_{(l)}^2(Q^2)
&=&
  \hat{m}_{(l)}^2\, \left(a_{s}(Q^2)\right)^{\nu_0}
  \left[1 + \sum_{m\geq 1}^{\infty} e^{(l)}_m\,
        \left(a_{s}(Q^2)\right)^m
  \right] \, ,
\label{eq:Z-2loop}
\end{eqnarray}
and setting $\mu^2=Q^2$, we find
\begin{equation}
  \left[3\,\hat{m}_b^2\right]_{(l)}^{-1}\,
  \widetilde{D}^{(l)}_{\text{S}}(Q^2)
 =
   \left(a^{(l)}_{s}(Q^2)\right)^{\nu_0}
  +\sum_{n\geq1}^{l}d_n\,
   \left(a^{(l)}_{s}(Q^2)\right)^{n+\nu_0}
 +\sum_{m\geq1}^{\infty}\Delta^{(l)}_m\,
   \left(a^{(l)}_{s}(Q^2)\right)^{m+\nu_0}
\label{eq:D-approx}
\end{equation}
with
\begin{equation}
  \Delta^{(l)}_m
= e_m^{(l)}
  + \sum_{ k\geq1}^{{\rm min}[l,m-1]}d_k\,e_{m-k}^{(l)}\, .
\label{eq:d-tild_1}
\end{equation}
Note that we have separated the mass-evolution multiloop effects
(third term of Eq.\ (\ref{eq:D-approx})) from the original series
expansion of $D$ (truncated at $n=l$).
These contributions are represented by the second term on the RHS of
Eq.\ (\ref{eq:D-approx}).
In practice, for $Q\geq 2$~GeV, i.e., for $\alpha_s\leq0.4$,
the truncation at $m=l+4$ of the summation in (\ref{eq:Z-2loop})
is sufficient, given that the truncation error is less than
1 percent.

By applying the analytization operation ${\bf A}_{\rm M}$ to the
quantity $\widetilde{D}_{\rm S}^{(l)}(Q^2)$, we finally obtain
\begin{eqnarray}
  \widetilde{R}_\text{S}^{(l)\text{MFAPT}}
& = &  \textbf{A}_\text{M}[D^{(l)}_{\text{S}}]
\nonumber \\
& = &
       3\,\hat{m}_{(l)}^2\,
       \left[{\mathfrak a}_{\nu_{0}}^{(l)}
             +\sum_{n\geq1}^{l} d_{n}^{}
              {\mathfrak a}_{n+\nu_{0}}^{(l)}
             +\sum_{m\geq1}^{}\Delta_{m}^{(l)}
              {\mathfrak a}_{m +\nu_{0}}^{(l)}
       \right]\ ,
\label{eq:R-MFAPT}
\end{eqnarray}
where we have used the shorthand notation
\begin{eqnarray}
  \left[a_s(s)^{\nu}\right]_\text{an}
=
  {\mathfrak a}_\nu^{(l)}(s)
\equiv
  \left(\frac{4}{b_0}\right)^{\nu}
  {\mathfrak A}_{\nu}^{(l)}(s) \ .
\label{eq:Che.Anal.Coupl}
\end{eqnarray}
This expression contains, by means of the coefficients
$\Delta_{n}^{(l)}~(~e^{(l)}_k)$
and the couplings
${\mathfrak a}_{n+\nu_{0}}^{(l)}$,
all renormalization-group terms contributing to this order.
On the other hand, the resummed $\pi^2$ terms are integral parts
of the analytic couplings
${\mathfrak a}_{m+\nu_{0}}^{(l)}$.

%
\begin{figure}[t]
\vspace{0.3cm}
\centerline{\includegraphics[width=0.45\textwidth]{%
 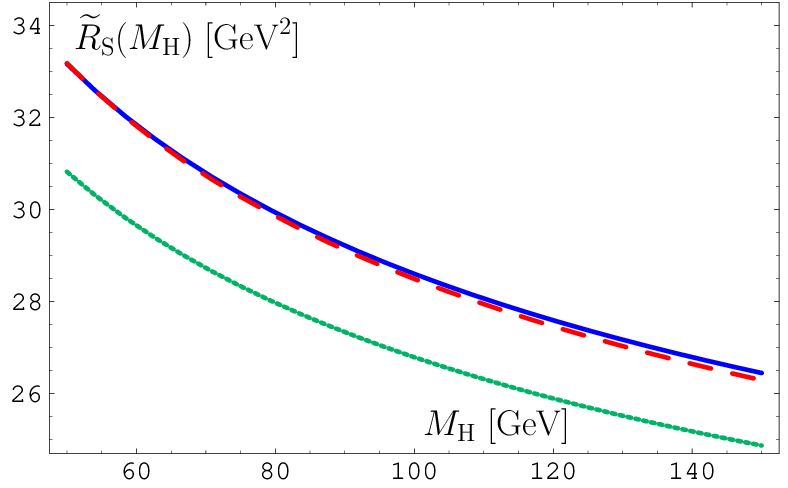}
 \vspace*{2mm}
 ~~~~~~\includegraphics[width=0.45\textwidth]{%
 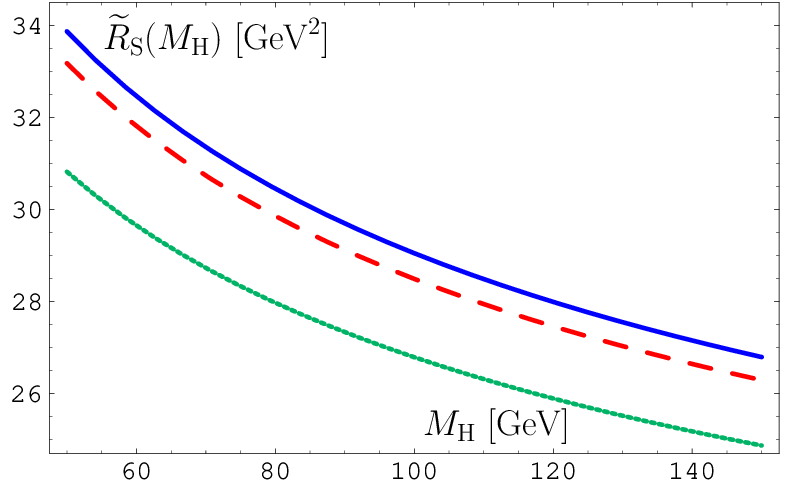}}
\vspace{0.2cm}
\caption{Predictions for the quantity
   $\widetilde{R}_{\rm S}(M^2_{\rm H})$ calculated in different
   perturbative approaches within the $\overline{\rm MS}$
   scheme:
   Standard perturbative QCD \protect\cite{ChKS97} at the
   loop level $l=4$ (dashed red
   line with $\Lambda_{N_f=5}=231$~MeV),
   Broadhurst-Kataev-Maxwell (``naive non-Abelianization'')
   including the
   $O(\left(a_{s}\right)^{\nu_0}  A_{4}(a_{s}))$-terms,
   \protect\cite{BKM01}---(dotted green line with
   $\Lambda_{N_f=5}=111$~MeV),
   and FAPT \protect\cite{BMS06} obtained from
   Eq.\ (\ref{eq:R-MFAPT}) for $N_f=5$ (solid blue
   line), displayed for two different loop orders:
   $l=2$ (left panel, $\Lambda_{N_f=5}=263$~MeV) and $l=3$ (right
   panel with $\Lambda_{N_f=5}=261$~MeV).
   The value of $\Lambda_{N_f=5}$~MeV in all cases corresponds to
   ${\mathfrak A}_1^{(1)}(s=m_Z^2;N_f=5)=0.120$.}
\label{fig:6}
\end{figure}

The results for
$\widetilde{R}_{\rm S}(M^2_{\rm H})$ as a function of the Higgs mass
$M_{\rm H}$, calculated with different perturbative approaches in the
$\overline{\rm MS}$ scheme, are shown in Fig.\ \ref{fig:6}.
The long-dashed curve shows the predictions obtained in \cite{ChKS97}
using standard fixed-order perturbative QCD at the $l=4$ loop level of
the expansion.
The solid curve next to it represents the FAPT prediction
(cf.\ (\ref{eq:R-MFAPT})), including in the second sum all
evolution effects up to $m=l+4$ and fixing the active flavor number
to $N_f=5$.
On the other hand, the contributions of the higher-loop
renormalization-group dependent terms are accumulated in the
coefficients $\Delta_{m}^{(l)}$ by means of the parameters
$\gamma_i$ and $b_j$.
Obviously, the standard perturbative QCD approach and FAPT yield
similar predictions for this observable, starting with the
two-loop running.
The reason for the slightly larger FAPT prediction lies in the fact
that the coefficients ${\mathfrak a}_{\nu}^{}$ contain by means of
the index $\nu_0$ the resummed contribution of an \textit{infinite}
series of $\pi^2$-terms (as well as all $\gamma_0$ and $b_0$ terms)
that renders them ultimately smaller than the corresponding powers
of the standard coupling.
Finally, the dotted green curve about 8\% below the other lines,
in both panels of Fig.\ \ref{fig:6}, illustrates the estimate obtained
with the ``naive non-Abelianization'' and an optimized power-series
expansion that makes use of the ``contour-improved integration
technique'' \cite{BKM01}.

%
\begin{figure}[t!]
 \centerline{\includegraphics[width=0.55\textwidth]{
  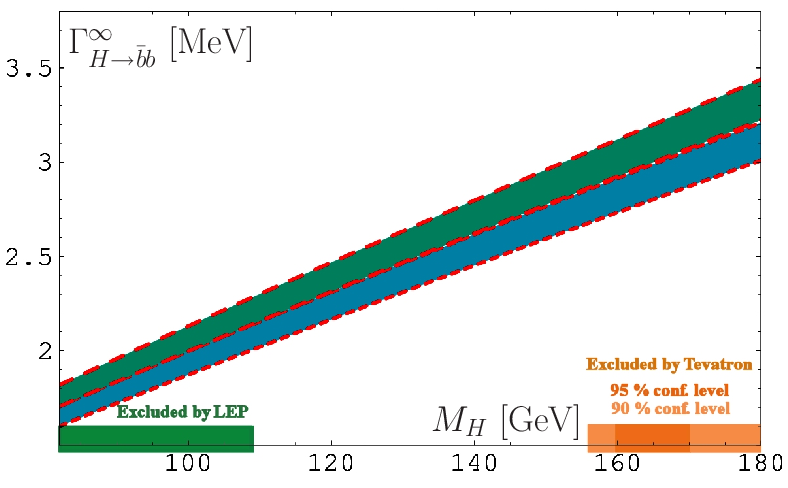}\vspace*{-1mm}}
  \caption{Predictions for the the two-loop width
    $\Gamma_{H \to b\bar{b}}^{\infty}$
    as a function of the Higgs-boson mass $M_{\rm H}$ using
    resummed FAPT \protect\cite{BMS10}.
    The lower strip is obtained by varying the mass in the interval
    $\hat{m}_{(2)}=8.22\pm0.13$~GeV according to the Penin--Steinhauser
    estimate
    $\overline{m}_b(\overline{m}_b^2)=4.35\pm0.07$~GeV
    \protect\cite{PeSt02},
    while the upper strip shows the corresponding one-loop result.
    The mass window of the Higgs-boson that is experimentally accessible
    is indicated.
    \label{fig:HiggsStrip2L}}
\end{figure}

The above Higgs decay results can be further extended by resumming the
whole nonpower series \cite{BMS10}, whereas the case of the three-loop
running coupling was elaborated in detail in~\cite{ABIP11strba}.
This way, one may extract more reliable predictions for the width
$\Gamma_{H\to b\bar{b}}$
in terms of the Higgs-boson mass $M_{\rm H}$.
This is indeed possible by employing an appropriate generating function
to determine the next two higher expansion coefficients reaching at the
truncation level of N$^3$LO an accuracy of the order of $1\%$, which
is even better than the 2\% uncertainty involved in the
$\overline{m}_b(\overline{m}_b^2)$ estimates.
The predictions for the one-loop and the two-loop width are illustrated
in Fig.\ \ref{fig:HiggsStrip2L} vs. the mass of the Higgs boson for
which the Penin--Steinhauser value
$\overline{m}_b(\overline{m}_b^2)=4.35\pm0.07$~GeV
\protect\cite{PeSt02} was used.
Adopting instead the RG-effective $b$-quark mass
$\overline{m}_b(\overline{m}_b^2)=4.19\pm0.05$~GeV,
calculated by K\"uhn and Steinhauser \cite{KuSt01},
one obtains results somewhat reduced by about $7\%$.

Very recently, the Atlas \cite{ATLAS2012} and the CMS \cite{CMS2012}
Collaborations from the Large Hadron Collider (LHC) at CERN have
reported the discovery of a new particle consistent with the Standrad
Model (SM) Higgs boson.
The mass measured by CMS is $M_{\rm H}=125.3$, while that found by
Atlas is slightly larger (around 126~GeV).
A confirmation at the five standard-deviation level of certainty
that this is the Higgs boson is still lacking.
Using a mass value of the SM Higgs boson around 125~GeV, we can extract
from Fig.\ \ref{fig:HiggsStrip2L} the following prediction
\begin{equation}
 \Gamma_{H\to b\bar{b}}=2.4\pm0.15~{\rm MeV} \, .
\label{eq:Higgs-gamma-pred}
\end{equation}
This estimate covers both the one loop result (upper strip) as well
as the two-loop one (lower strip) and is within the range of
$BR(H\to b\bar{b})$ measured very recently by the D0 Collaboration
\cite{D02012_Higgs}.

\section{Conclusions}
\label{sec:concl}
FAPT derives from APT and generalizes it both conceptually and
technically.
Conceptually, it extends the analyticity requirement from the running
coupling and its powers to the hadronic amplitude as a whole.
Technically, it extends the formalism from integer powers of the
coupling to real ones in the spacelike as well as in the timelike
region.
More recently, progress has been achieved by endowing FAPT with a
varying flavor number across heavy-quark thresholds:
``flavor-corrected global FAPT''
(see \cite{AB08gfapt} for explanations and details).
Its numerical realization has quite recently been implemented
in the form of the \texttt{Mathematica} package
\texttt{FAPT.m}~\cite{BaKha12}.
Yet FAPT is not a complete formalism.
We still do not really know how to consider Sudakov
gluon resummation within FAPT because under the analytization
requirement resummation does not mean exponentiation.
This important issue has yet to be understood and developed.
Also the application of FAPT to compute power corrections to the
spacelike pion's LO electromagnetic form factor and the inclusive
Drell--Yan cross section, started in \cite{KS01}, has so far not
been exploited beyond the leading power and for other hadronic
quantities.

\begin{acknowledgments}
I wish to thank A.\ P.\ Bakulev and S.\ V.\ Mikhailov for a fruitful
collaboration and for help in preparing the updated version of
this draft.
I am also indebted to A.\ L.\ Kataev for useful remarks.
This work was partially supported by the Heisenberg-Landau
Program (Grant 2009) and the Deutsche Forschungsgemeinschaft under
contract 436RUS113/881/0.
\end{acknowledgments}


\end{document}